\documentclass[11pt]{article}
\usepackage{graphics,epsfig}
\usepackage[latin1]{inputenc}
\usepackage[english,activeacute]{babel}
\usepackage[]{graphicx}
\usepackage{latexsym}
\usepackage{amsmath, amsthm, amsfonts, amssymb}
\usepackage{verbatim}

\newcommand{\nd}{\noindent}

\begin{document}

\newtheorem{theo}{Theorem}[section]
\newtheorem{definition}[theo]{Definition}
\newtheorem{lem}[theo]{Lemma}
\newtheorem{prop}[theo]{Proposition}
\newtheorem{coro}[theo]{Corollary}
\newtheorem{exam}[theo]{Example}
\newtheorem{rema}[theo]{Remark}
\newtheorem{example}[theo]{Example}
\newtheorem{principle}[theo]{Principle}
\newcommand{\ninv}{\mathord{\sim}}
\newtheorem{axiom}[theo]{Axiom}

\title{A discussion on the origin of quantum probabilities}

\author{{\sc Federico Holik}$^{1,2}$ \ {\sc ,} \ {\sc Angel Plastino}$^{3}$ \ {\sc and} {\sc Manuel S\'{a}enz}$^{2}$}

\maketitle

\begin{center}

\begin{small}
1- Universidad Nacional de La Plata, Instituto
de F\'{\i}sica (IFLP-CCT-CONICET), C.C. 727, 1900 La Plata, Argentina \\
2- Departamento de Matem\'{a}tica - Facultad de Ciencias Exactas y
Naturales\\ Universidad de Buenos Aires - Pabell\'{o}n I, Ciudad
Universitaria \\ Buenos Aires, Argentina.\\
3- Universitat de les Illes Balears and IFISC-CSIC, 07122 Palma de
Mallorca, Spain \\
\end{small}
\end{center}

\vspace{1cm}

\begin{abstract}
\noindent We study the origin of quantum probabilities as arising
from non-boolean propositional-operational structures. We apply
the method developed by Cox to non distributive lattices and develop an alternative formulation of
non-Kolmogorvian probability measures for quantum
mechanics. By generalizing the method presented in previous works,
we outline a general framework for the deduction of probabilities in
general propositional structures represented by lattices
(including the non-distributive case).
\end{abstract}
\bigskip
\noindent

\begin{small}
\centerline{\em Key words: Quantum Probability-Lattice
theory-Information theory}
\end{small}

\bibliography{pom}

\begin{thebibliography}{00}

\bibitem{BvN} G. Birkhoff and J. von Neumann, Annals Math. {\bf 37} (1936)
823-843.

\bibitem{uno} Dov M. Gabbay, John Woods, {\it The Many Valued and Nonmonotonic Turn in Logic} (Elsevier, Amsterdam, 2007), p.
205.

\bibitem{RedeiHandbook} M. Redei ``The Birkhoff-von Neumann Concept of quantum Logic'', in Handbook of Quantum Logic and Quantum Structures, K. Engesser, D. M. Gabbay and D. Lehmann, eds., Elsevier (2009).

\bibitem{Gudder-StatisticalMethods} S. P. Gudder, \textit{Stochastic
Methods in Quantum Mechanics} North Holland, New York - Oxford
(1979).

\bibitem{gudderlibro78} S. P. Gudder, in {\it Mathematical Foundations
of Quantum Theory}, A. R. Marlow, ed., Academic, New York, (1978).

\bibitem{dallachiaragiuntinilibro} M. L. Dalla Chiara, R. Giuntini,
and R. Greechie, {\it Reasoning in Quantum Theory}, Kluwer Acad.
Pub., Dordrecht, (2004).

\bibitem{mikloredeilibro} M. R\'{e}dei, \textit{Quantum Logic in Algebraic
Approach}, Kluwer Academic Publishers, Dordrecht, (1998).

\bibitem{Redei-Summers2006} M. R\'{e}dei and S. Summers, \textit{Studies in History and Philosophy of Science Part B: Studies in
History and Philosophy of Modern Physics} Volume \textbf{38}, Issue
\textbf{2}, (2007) 390-417.

\bibitem{mackey-book} G. Mackey \textit{Mathematical foundations of quantum mechanics} New York:
W. A. Benjamin (1963).

\bibitem{Davies-Lewies} E. Davies and J. Lewis, \textit{Commun. Math. Phys.} \textbf{17},
(1970) 239-260.

\bibitem{Srinivas} M. Srinivas, \textit{J. Math. Phys.} \textbf{16}, (1975)
1672.

\bibitem{AcacioSuppes} J. Acacio de Barros and P. Suppes,
arXiv:quant-ph/0001017v1 (2000).

\bibitem{Anastopoulos} C. Anastopoulos, \textit{Annals Of Physics} \textbf{313},
(2004) 368-382.

\bibitem{Rau} J. Rau, \textit{Annals Of Physics} \textbf{324}, (2009) 2622-2637.

\bibitem{KolmogorovProbability} Kolmogorov, A.N. Foundations of Probability Theory; Julius
Springer: Berlin, Germany, (1933).

\bibitem{CoxPaper} Cox, R.T. Probability, frequency, and reasonable expectation. \textit{Am.
J. Phys.} \textbf{14}, (1946) 1-13.

\bibitem{CoxLibro} Cox, R.T. \textit{The Algebra of Probable Inference}; The Johns Hopkins
Press: Baltimore, MD, USA, (1961).

\bibitem{Boole} Boole, G. \textit{An Investigation of the Laws of Thought},
Macmillan: London, UK, (1854).

\bibitem{Knuth-2004a} Knuth, K.H. Deriving laws from ordering relations. In \textit{Bayesian
Inference and Maximum Entropy Methods in Science and Engineering},
Proceedings of 23rd International Workshop on Bayesian Inference and
Maximum Entropy Methods in Science and Engineering; Erickson, G.J.,
Zhai, Y., Eds.; American Institute of Physics: New York, NY, USA,
pp. 204-235 (2004).

\bibitem{Knuth-2004b} Knuth, K.H. Measuring on lattices. In \textit{Bayesian Inference and
Maximum Entropy Methods in Science and Engineering}, Proceedings of
23rd International Workshop on Bayesian Inference and Maximum
Entropy Methods in Science and Engineering; Goggans, P., Chan, C.Y.,
Eds.; American Institute of Physics: New York, NY, USA,; Volume 707,
pp. 132-144 (2004).

\bibitem{Knuth-2005a} Knuth, K.H. Valuations on lattices and their application to
information theory. In Proceedings of the 2006 IEEE World Congress
on Computational Intelligence, Vancouver, Canada, July (2006).

\bibitem{Jaynes-1957a} E. T. Jaynes, \textit{Phys. Rev.} Vol. \textbf{106}, Number 4
(1957).

\bibitem{Jaynes-1957b} E. T. Jaynes, \textit{Phys. Rev.} Vol. \textbf{108}, Number 2
(1957).

\bibitem{Caticha-99} A. Caticha, \textit{Phys. Rev. A} \textbf{57} (1998) 1572-1582.

\bibitem{Knuth-2005b} Knuth, K.H. Lattice duality: The origin of probability and
entropy. \textit{Neurocomputing} 67C, (2005) 245-274.

\bibitem{Symmetry} P. Goyal and K. Knuth, \textit{Symmetry} \textbf{3} (2),
(2011) 171-206, doi:10.3390/sym3020171.

\bibitem{GoyalKnuthSkilling} Goyal, P., Knuth, K.H. and Skilling, J., \textit{Phys. Rev. A} \textbf{81} (2010) 022109.

\bibitem{KnuthSkilling-2012} Knuth, K.H. and Skilling,
arXiv:1008.4831v1 (2012).

\bibitem{mackey57} G. W. Mackey, \textit{Amer. Math. Monthly}, Supplement
{\bf 64} (1957) 45-57.

\bibitem{jauch} J. M. Jauch, {\it Foundations of Quantum Mechanics}, Addison-Wesley, Cambridge, (1968).

\bibitem{piron} C. Piron, {\it Foundations of Quantum Physics}, Addison-Wesley, Cambridge,
(1976).

\bibitem{kalm83} G. Kalmbach, {\it Orthomodular Lattices}, Academic
Press, San Diego, (1983).

\bibitem{kalm86} G. Kalmbach, {\it Measures and Hilbert Lattices}, World Scientific, Singapore, (1986).

\bibitem{vadar68} V. Varadarajan, {\it Geometry of Quantum Theory I}, van Nostrand, Princeton, (1968).

\bibitem{vadar70} V. Varadarajan, {\it Geometry of Quantum Theory
II}, van Nostrand, Princeton, (1970).

\bibitem{greechie81} J. R. Greechie, in {\it Current Issues in Quantum
Logic}, E. Beltrameti and B. van Fraassen, eds., Plenum, New York,
(1981) pp. 375-380.

\bibitem{giunt91} R. Giuntini, {\it Quantum Logic and Hidden
Variables}, BI Wissenschaftsverlag, Mannheim, (1991).

\bibitem{pp91} P. Pt\'{a}k and S. Pulmannova, {\it Orthomodular
Structures as Quantum Logics}, Kluwer Academic Publishers,
Dordrecht, (1991).

\bibitem{belcas81} E. G. Beltrametti and G. Cassinelli, {\it The
Logic of Quantum Mechanics}, Addison-Wesley, Reading, (1981).

\bibitem{dvupulmlibro} A Dvure\v{c}enskij and S. Pulmannov\'{a},
{\it New Trends in Quantum Structures}, Kluwer Acad. Pub.,
Dordrecht, (2000).

\bibitem{HandbookofQL} \textit{Handbook Of Quantum Logic And Quantum Structures}
(Quantum Logic), Edited by K. Engesser, D. M. Gabbay and D. Lehmann,
North-Holland (2009).

\bibitem{aertsdaub1} D. Aerts and I. Daubechies, \textit{Lett. Math. Phys.}
{\bf 3} (1979) 11-17.

\bibitem{aertsdaub2} D. Aerts and I. Daubechies, \textit{Lett. Math. Phys.}
{\bf 3} (1979) 19-27.

\bibitem{aertsjmp83} D. Aerts, \textit{J. Math. Phys.} {\bf 24} (1983) 2441.

\bibitem{aertsrepmathphys84a} D. Aerts, \textit{Rep. Math. Phys} {\bf 20}
(1984) 421-428.

\bibitem{aertsjmp84b} D. Aerts, \textit{J. Math. Phys.} {\bf 25} (1984)
1434-1441.

\bibitem {Mielnik-68} B. Mielnik, \textit{Commun. Math. Phys.} \textbf{9} (1968)
55-80.

\bibitem {Mielnik-69} B. Mielnik, \textit{Commun. Math. Phys.} \textbf{15} (1969)
1-46.

\bibitem {Mielnik-74} B. Mielnik, \textit{Commun. Math. Phys.} \textbf{37} (1974)
221-256.

\bibitem{Ludwig1} G. Ludwig, \textit{Commun. Math. Phys.} \textbf{4}, (1967) 331-348.

\bibitem{Ludwig2} G. Ludwig, \textit{Commun. Math. Phys.} \textbf{9}, (1968) 1-12.

\bibitem{vN} J. von Neumann, {\it Mathematical Foundations of Quantum
Mechanics}, Princeton University Press, 12th. edition, Princeton,
(1996).

\bibitem{ReedSimon} M. Reed and B. Simon, \textit{Methods of modern
mathematical physics} I: Functional analysis, Academic Press, New
York-San Francisco-London (1972).

\bibitem{Soler-1995} M. Sol\`{e}r, \textit{Communications in Algebra}
\textbf{23}, (1995) 219-243.

\bibitem{wilce} A. Wilce, {\it Quantum Logic and Probability Theory},
The Stanford Encyclopedia of Philosophy (Spring 2009 Edition),
Edward N. Zalta (ed.), URL =
http://plato.stanford.edu/archives/spr2009/entries/qt-quantlog/.
Archive edition: Spring 2009.

\bibitem{Barnum-Wilce-2006} H. Barnum, J. Barret, M. Leifer and A.
Wilce, Phys. Rev. Lett. \textbf{99}, 240501 (2007).

\bibitem{Barnum-Wilce-2009} H. Barnum and A. Wilce, arXiv:0908.2352v1
[quant-ph] (2009); Electronic Notes in Theoretical Computer Science
Volume \textbf{270}, Issue 1, Pages 3-15,(2011).

\bibitem{Barnum-Wilce-2010} H. Barnum, R. Duncan and A Wilce, arXiv:1004.2920v1
[quant-ph] (2010).

\bibitem{Redei99} M. R\'{e}dei, \textit{The Mathematical Intelligencer},
\textbf{21} (4), (1999) 7-12.

\bibitem{aczel-book} Aczél, J., \textit{Lectures on Functional Equations and Their
Applications}, Academic Press, New York, (1966).

\bibitem{Gleason} A. Gleason, J. Math. Mech. \textbf{6}, (1957) 885-893.

\bibitem{Gleason-Dvurechenski-2009} D. Buhagiar, E. Chetcuti and A.
Dvure\v{c}enskij, Found. Phys. \textbf{39}, 550-558 (2009).

\bibitem{chom} N. Chomsky, IRE Transactions on Information Theory {\bf 2}, 113
(1956).

\bibitem{svozillibro} K. Svozil, {\it Quantum Logic}, Springer-Verlag, Singapore, (1998).

\end{thebibliography}

\section{Introduction}

\noindent Quantum probabilities\footnote{By the term ``quantum
probabilities", we mean the probabilities that appear in quantum
theory. As is well known, they are ruled by the well known formula
$\mbox{tr}(\rho P)$, where $\rho$ is a density matrix representing
a general quantum state and $P$ is a projection operator
representing an event (see Section \ref{s:probabilities} of this
work for details).} posed an intriguing question from the very
beginning of quantum theory. It was rapidly realized that
probability amplitudes of quantum process obeyed rules of a non
classical nature, as for example, the sum rule of probability
amplitudes giving rise to interference terms or the nonexistence
of joint distributions for noncommuting observables. In 1936 von
Neumann wrote the first work ever to introduce quantum logics
\cite{BvN,uno,RedeiHandbook}, suggesting that quantum mechanics
requires a propositional calculus substantially different from all
classical logics. He rigorously isolated a new algebraic structure
for quantum logics, and studied its connections with quantum
probabilities. Quantum and classical probabilities have points in
common as well as differences. These differences and the
properties of quantum probabilities have been intensively studied
in the literature
\cite{Gudder-StatisticalMethods,gudderlibro78,dallachiaragiuntinilibro,mikloredeilibro,Redei-Summers2006,mackey-book,Davies-Lewies,Srinivas,AcacioSuppes,Anastopoulos,Rau}.
It is important to remark that not all authors believe that
quantum probabilities are essentially of a different nature than
those which arise in probability theory (see for example
\cite{Symmetry} for a recent account). Thought this is a major
question for probability theory and physics, it is not our aim in
this work to settle this discussion.

\noindent There exist two important axiomatizations of classical
probabilities. One of them was provided by Kolmogorov
\cite{KolmogorovProbability}, a set theoretical approach based on
boolean sigma algebras of a sample space. Probabilities are defined
as measures over subsets of a given set. Thus, the Kolmogorovian
approach is set theoretical and usually identified (but not
necessarily) with a frequentistic interpretation of probabilities.
Some time later it was realized that quantum probabilities can be
formulated as measures over non boolean structures (instead of
boolean sigma algebras). This is the origin of the name
``non-boolean or non-kolmogorovian'' probabilities
\cite{Redei-Summers2006}. It is remarkable that the creation of
quantum theory and the works on the foundations of probability by
Kolmogorov where both developed at the same time, in the twenties.

\noindent An alternative approach to the Kolmogorovian construction
of probabilities was developed by R. T. Cox
\cite{CoxPaper,CoxLibro}. Cox starts with a propositional calculus,
intended to represent assertions which portray our knowledge about
the world or system under investigation. As it is well known since
the work of Boole \cite{Boole}, propositions of classical logic (CL)
can be represented as a Boolean lattice, i.e., an algebraic
structure endowed with lattice operations ``$\wedge$", ``$\vee$",
and ``$\neg$", which are intended to represent conjunction,
disjunction, and negation, respectively, together with a partial
order relation ``$\leq$" which is intended to represent logical
implication. Boolean lattices (as seen from an algebraic point of
view) can be characterized by axioms
\cite{Knuth-2004a,Knuth-2004b,Knuth-2005a}. By considering
probabilities as an inferential calculus on a boolean lattice, Cox
showed that the axioms of classical probability can be deduced as a
consequence of lattice symmetries, using entropy as a measure of
information. Thus, differently form the set theoretical approach of
Kolmogorov, the approach by Cox considers probabilities as an
inferential calculus.

\vskip 3mm  \nd It was recently shown that Feymann's rules of
quantum mechanics can be deduced from operational lattice structures
using a variant of Cox method
\cite{Caticha-99,Knuth-2005b,Symmetry,GoyalKnuthSkilling,KnuthSkilling-2012}
(see also \cite{Knuth-2004a,Knuth-2004b}). For example, in
\cite{Symmetry,GoyalKnuthSkilling} this is done by:

\begin{itemize}
\item first defining an operational propositional calculus on a
quantum system under study, and after that,
\item postulating that any quantum process (interpreted as a proposition
in the operational propositional calculus) can be represented by a
pair of real numbers and,
\item  using a variant of the method developed by Cox,
showing that these pairs of real numbers obey the sum and product
rules of complex numbers, and can then be interpreted as the
quantum probability amplitudes which appear in Feymann's rules.
\end{itemize}

\nd There is a long tradition with regards to  the application of
lattice theory to physics and many other disciplines. The quantum logical (QL) approach to quantum
theory (and physics in general), initiated by von Neumann in
\cite{BvN}, has been a traditional tool for studies on the
foundations of quantum mechanics (see for example
\cite{mackey57,jauch,piron,kalm83,kalm86,vadar68,
vadar70,greechie81,giunt91,pp91,belcas81,gudderlibro78}, and for a
complete bibliography \cite{dallachiaragiuntinilibro},
\cite{dvupulmlibro}, and \cite{HandbookofQL}).

\nd The (QL) approach to physics bases itself on defining elementary
tests and propositions for physical systems and then, studying the
nature of these propositional structures. In some approaches,
this is done in an operational way
\cite{piron,aertsdaub1,aertsdaub2,aertsjmp83,aertsrepmathphys84a,aertsjmp84b},
and is susceptible of considerable generalization to arbitrary
physical systems (not necessarily quantum ones). That is why the
approach is also called \emph{operational quantum logic}
(OQL)\footnote{In this paper we will use the terms QL and OQL
interchangeably, but it is important to remark that -though similar-
they are different approaches.}. One of the most important goals of
OQL is to impose operationally motivated axioms on a lattice
structure in order that it can be made isomorphic to a projection
lattice on a Hilbert space. There are different positions in the
literature about the question of whether this goal has been achieved
or not \cite{Rau}, and also, of course, alternative operational
approaches to physics, as the \emph{convex operational} one
\cite{Mielnik-68,Mielnik-69,Mielnik-74,Ludwig1,Ludwig2,Gudder-StatisticalMethods}.
In this work, we are interested in the great generality of the OQL
approach. The operational approach presented in
\cite{Gudder-StatisticalMethods} bases itself only in the convex
formulation of any statistical theory, and it can be shown that the
more general structure which appears under reasonable operational
considerations is a \emph{$\sigma$-orthocomplemented orthomodular
poset}, a more general class than \emph{orthomodular lattices} (the
ones which appear in quantum theory). We will come back to these
issues and review the definitions for these structures below.

\nd  \fbox{\parbox{0.97\linewidth}{ In this work we complement the
work presented in \cite{Knuth-2004b,Knuth-2005a} asking the
following questions:

\begin{itemize}

\item is it possible to generalize Cox's method to arbitrary
lattices or more general algebraic structures?

\item what happens if the Cox's method mentioned above is applied to
general lattices (not necessarily distributive), representing
general physical systems? And in particular, what happens if it is
applied to the von Newmann's lattice of projection operators?

\item does the logical underlying structure of the theory determine
the form and properties of the probabilities?

\end{itemize}

\nd As we shall see below, it is possible to use these questions
to give an alternative formulation of quantum probabilities. We will show that once the
operational structure of the theory is fixed, the general
properties of probability theory are -in a certain sense to be clarified below- determined. We also discuss
the implications of our derivation for the foundations of quantum
physics and probability theory, and compare with ours  different
approaches: the one presented in \cite{Symmetry,GoyalKnuthSkilling}, the OQL
approach, the operational approach of
\cite{Gudder-StatisticalMethods}, and the traditional one
(represented by the von Neumann formalism of Hilbertian quantum
mechanics \cite{vN}).

\nd The approach presented here shows itself to be susceptible of
great generalization: we provide an algorithm for developing
generalized probabilities using a combination of the Cox's method
with the OQL approach. This opens the door to the development of
more general probability and information measures. This
methodology is  advantageous because, in the particular case of
quantum mechanics, it includes mixed states in a natural way, unlike other
approaches based only on pure states (like the ones presented in
\cite{KnuthSkilling-2012} and \cite{Caticha-99}).}}

\nd The paper is organized as follows. In Section
\ref{s:LogicOperational} we review the $QL$ approach to physics as
well as lattice theory. In Section \ref{s:CoxReview} we revisit
Kolmogorov's and Cox' approaches to probability. After that, in
Section \ref{s:probabilities}, we give a sketch concerning
quantum probabilities and their differences with  classical ones.
In Section \ref{s:KnuthGoyal} we discuss the approach developed in
\cite{Caticha-99}, \cite{Knuth-2005b}, \cite{Symmetry},
\cite{GoyalKnuthSkilling}, and \cite{KnuthSkilling-2012}. In
Section \ref{s:CoxNon-Boolean} we apply Cox's method for the formulation of
non-Kolmogorovian probabilities using the algebraic properties of non-boolean lattices and
study several examples. Finally, in section \ref{s:Conclusions}
some conclusions are drawn.

\section{The lattice/operational approach to
physics}\label{s:LogicOperational}

\nd The quantum logical approach to physics is vast and includes
different programs. We will concentrate on the path followed by von
Neumann and the operational approach developed by Jauch, Piron, and
others. First, we recall the relationship between projection
operators and elementary tests in $QM$. After studying the examples
of lattices applied to QM and CM, we review the main features of the $QL$ approach. The
reader familiar with these topics can skip this Section.

\subsection{Elementary notions of lattice theory}\label{s:LatticeTheory}

\noindent A partially ordered set (also called a poset) is a set
$X$ endowed with a partial ordering relation ``$<$"  satisfying

\begin{itemize}

\item 1- For all $x,y\in X$, $x<y$ and $y<x$ entail $x=y$

\item 2- For all $x,y,z\in X$, if $x<y$ and $y<z$, then $x<z$

\end{itemize}

\nd The notation ``$x\leq y$" is used to denote ``$x<y$" or
``$x=y$". A lattice $\mathcal{L}$ will be a poset in which any two
elements $a$ and $b$ have a unique supremum (the elements' least
upper bound ``$a\vee b$"; called their join) and an infimum
(greatest lower bound ``$a\wedge b$"; called their meet). Lattices
can also be characterized as algebraic structures satisfying
certain axiomatic identities imposed on operations ``$\vee$" and
``$\wedge$".  For a {\it complete} lattice all its subsets have
both a supremum (join) and an infimum (meet).

\vskip 3mm \nd A {\it bounded} lattice has a greatest (or maximum)
and least (or minimum) element, denoted $1$ and $0$ by convention
(also called top and bottom, respectively). Any lattice can be
converted into a bounded lattice by adding a greatest and least
element, and every non-empty finite lattice is bounded. For any set
$A$, the collection of all subsets of $A$ (called the power set of
$A$) can be ordered via subset inclusion to obtain a lattice bounded
by $A$ itself and the null set. Set intersection and union represent
the operations meet and join, respectively.

\vskip 3mm \nd  Every complete lattice is a bounded lattice. While
bounded lattice homomorphisms in general preserve only finite
joins and meets, complete lattice homomorphisms are required to
preserve arbitrary joins and meets. If $P$ is a bounded poset, an
orthocomplementation in $P$ a unary operation ``$\neg(\ldots)$" such that:

\begin{subequations}\label{e:ComplementationAxioms}
\begin{equation}\label{e:Complement1}
\neg(\neg(a)))=a
\end{equation}
\begin{equation}\label{e:Complement2}
a\leq b \longrightarrow \neg b\leq \neg a
\end{equation}
$a\vee \neg a$ and $a\wedge \neg a$ exist and both
\begin{equation}\label{e:Complement3}
a\vee \neg a=\mathbf{1}
\end{equation}
\begin{equation}\label{e:Complement4}
a\wedge \neg a=\mathbf{0}
\end{equation}
\end{subequations}

\nd hold. A bounded poset with ortocomplementation will be called an
orthoposet. An ortholattice, will be an orthoposet which is also a
lattice. For $a,\,b \in \mathcal{L}$ (an ortholattice or
orthoposet), we say that $a$ is orthogonal to $b$ ($ a \bot b$) iff
$a\le \neg b$.

\vskip 3mm \nd Distributive lattices are lattices for which the
operations of join and meet are distributed over each other. A
complete complemented lattice that is also distributive is a
Boolean algebra. For a distributive lattice, the complement of
$x$, when it exists, is unique. The prototypical examples of
Boolean algebras are collections of sets for which the lattice
operations can be given by set union and intersection, and lattice
complementation by set theoretical complementation.

\nd A {\it modular} lattice is one that satisfies the following
self-dual condition (\emph{modular law} or \emph{modular identity})

\begin{equation}\label{e:ModularIdentity}
x \leq b \longrightarrow x \vee (a \wedge b) = (x \vee a) \wedge b
\end{equation}

\nd Modular lattices arise naturally in algebra and in many other
areas of mathematics. For example, the subspaces of a finite
dimensional vector space form a modular lattice. Every distributive
lattice is modular. In a not necessarily modular lattice, there may
still be elements $b$ for which the modular law holds in connection
with arbitrary elements $a$ and $x$ ($\le b$). Such an element is
called a modular element. Even more generally, the modular law may
hold for a fixed pair $(a,b)$. Such a pair is called a modular pair,
and there are various generalizations of modularity related to this
notion and to semi-modularity. \vskip 3mm

\nd An orthomodular lattice will be an ortholattice satisfying the
orthomodular law:

\begin{equation}\label{e:ModularIdentity2}
x \leq b \longrightarrow x \vee (\neg x \wedge b) = b
\end{equation}

\vskip 3mm

\nd Orthomodularity is a weakening of modularity. As an example, the lattice
${\mathcal{L}}_{v\mathcal{N}}({\mathcal{H}})$ of closed subspaces of a Hilbert space $H$ (see Section \ref{s:ElementaryTests})
is orthomodular. ${\mathcal{L}}_{v\mathcal{N}}({\mathcal{H}})$ is modular only if $H$ is finite dimensional and strictly orthomodular
for the infinite dimensional case.

\vskip 3mm

\nd The concept of  lattice's atom is of great physical importance.
If $\mathcal{L}$ has a least element $ 0$, then an element $x$ of
$\mathcal{L}$ is an {\it atom}  if $0 < x$ and there exists no
element $y$ of $\mathcal{L}$ such that $0 < y < x$. One  says that
$\mathcal{L}$ is: \newline i) {\it Atomic}, if for every nonzero
element $x$ of $\mathcal{L}$, there exists an atom $a$ of
$\mathcal{L}$ such that $ a \leq x$
\newline ii) Atomistic, if every element of $\mathcal{L}$ is a
supremum of atoms.

\subsection{Elementary measurements and projection
operators}\label{s:ElementaryTests}

\nd In QM, an elementary measurement given by a yes-no experiment
(i.e., a test in which we get the answer ``yes" or the answer
``no"), is represented by a projection operator. If $\mathbb{R}$ is
the real line, let $B(\mathbb{R})$ be the family of subsets of
$\mathbb{R}$ such that

\begin{itemize}
\item 1 - The family is closed under set theoretical complements.

\item 2 - The family is closed under denumerable unions.

\item 3 - The family includes all open intervals.
\end{itemize}

\noindent The elements of $B(\mathbb{R})$ are the \emph{Borel
subsets} of $\mathbb{R}$ \cite{ReedSimon}. Let $\mathcal{P}(\mathcal{H})$ be
the set of all projection operators (or equivalently, the set of closed subspaces of $\mathcal{H}$).
In QM, a projection valued measure (PVM) $M$, is a mapping

\begin{subequations}

\begin{equation}
M: B(\mathbb{R})\rightarrow \mathcal{P}(\mathcal{H})
\end{equation}

\noindent such that

\begin{equation}
M(\emptyset)=0
\end{equation}
\begin{equation}
M(\mathbb{R})=\mathbf{1}
\end{equation}
\begin{equation}
M(\cup_{j}(B_{j}))=\sum_{j}M(B_{j}),\,\,
\end{equation}

\noindent for any disjoint denumerable family ${B_{j}}$. Also,

\begin{equation}
M(B^{c})=\mathbf{1}-M(B)=(M(B))^{\bot}
\end{equation}

\end{subequations}

\nd Any elementary measurement is represented by a projection operator \cite{vN}.
All operators representing observables can be expressed in
terms of PVM's (and so, reduced to sets of elementary
measurements), via the spectral decomposition theorem, which
asserts that the set of spectral measurements may be put in a
bijective correspondence with the set $\mathcal{A}$ of self
adjoint operators of $\mathcal{H}$ \cite{ReedSimon}.

\vskip 3mm  \nd The set of closed subspaces $\mathcal{P}(\mathcal{H})$ of any quantum system can be endowed with a lattice structure:
$\mathcal{L}_{v\mathcal{N}}({\mathcal{H}})=
<\mathcal{P}(\mathcal{H}),\ \leq ,\ \wedge,\ \vee,\ \neg,\ 0,\ 1>$, where ``$\leq$'' is the set theoretical inclusion ``$\subseteq$'', ``$\wedge$'' is set theoretical intersection ``$\cap$'',
``$\vee$'' is the closure of the sum``$\oplus$'',
$0$ is the empty set $\emptyset$, $1$ is the total space
$\mathcal{H}$ and $\neg(S)$ is
the orthogonal complement of a subspace $S$
\cite{mikloredeilibro}. Closed subspaces can be put in one to one
correspondence with projection operators. \emph{Thus, elementary
tests in QM, which are represented by projection operators, can be
endowed with a lattice structure}. This lattice was called
``Quantum Logic" by Birkhoff and von Neumann \cite{BvN}. We will
refer to this lattice as the \emph{von Neumann-lattice
($\mathcal{L}_{v\mathcal{N}}(\mathcal{H})$})
\cite{mikloredeilibro}.

\vskip 3mm

\nd The analogous of this structure in Classical Mechanics ($CM$)
was provided by Birkoff and von Neumann \cite{BvN}. Take for
example the following operational propositions on a classical
harmonic oscillator: ``the energy is equal to $E_0$" and ``the
energy is lesser or equal than $E_0$". The first one corresponds
to an ellipse in phase space, and the second to the ellipse and
its interior. This simple example shows that operational
propositions in $CM$ can be represented by subsets of the  phase
space. Thus, given a classical system $S$ with phase space
$\Gamma$, let $\mathcal{P}(\Gamma)$ represent the set formed by
all the subsets of $\Gamma$. This set can be endowed with a lattice
structure as follows. If ``$\vee$" is represented by set union,
``$\wedge$" by set intersection, ``$\neg$" by set complement (with
respect to $\Gamma$), $\leq$ is represented by set inclusion, and $\textbf{0}$ and $\textbf{1}$ are
represented by $\emptyset$ and $\Gamma$ respectively, then
$<\mathcal{P}(\Gamma),\ \leq,\ \wedge,\ \vee,\ \neg,\ 0,\ 1>$ conform a
complete bounded lattice. This is the lattice of propositions of a
classical system, which as it is well known, \emph{is a boolean
one}. Thus, $\mathcal{P}(\Gamma)$, as well as
$\mathcal{P}(\mathcal{H})$, can be endowed with a propositional lattice structure.

\subsection{The Quantum Logical Approach to Physics}

\nd We have seen that operational propositions of quantum and
classical systems can be endowed with lattice structures. These
lattices where boolean for classical systems, and non distributive
for quantum ones. This fact, discovered by von Neumann \cite{BvN},
raised a lot of interesting questions. The first one is: is it
possible to obtain the formalism of $QM$ (as well as $CM$) by
imposing suitable axioms on a lattice structure? The surprising
answer is \emph{yes, it is possible}. But the road which led to
this result was fairly difficult and full of obstacles. In the
first place, it was a very difficult mathematical task to
demonstrate that a suitably chosen set of axioms on a lattice
would yield a representation theorem which would allow one to
recover Hilbertian $QM$. The first result was obtained by Piron,
and the final demonstration was given by Sol\`{e}r in 1995
\cite{Soler-1995} (see also \cite{dallachiaragiuntinilibro}, page
72). One of the  advantages ascribed to this approach was that the
axioms imposed on a lattice structure could be given a clear
operational interpretation: unlike the Hilbert space formulation,
whose axioms have the disadvantage of being \emph{ad hoc} and
physically unmotivated, the quantum logical approach would be
clearer and more intuitive from a physical point of view. But of
course, the operational validity of the axioms imposed on the
lattice structure was criticized by many authors (as an example,
see \cite{Rau}).

\vskip 3mm

\nd The second important question raised by the von Neumann
discovery was: given that $QM$ and $CM$ can be described by
operational lattices, is it possible to formulate the entire
apparatus of physics in lattice theoretical terms? Given
\emph{any} physical system, quantum, classical, or obeying more
general tenets, it is always possible to define an operational
propositional structure on it using the notion of elementary
tests. A very general approach to physics can be given using
\emph{event structures}, which are sets of events endowed with
probability measures satisfying certain axioms
\cite{Gudder-StatisticalMethods}. It can be shown (see
\cite{Gudder-StatisticalMethods}, Chapter 3) that any event
structure is isomorphic to a \emph{$\sigma$-orthocomplete
orthomodular poset}, which is an orthocomplemented poset
$\mathcal{P}$, satisfying the orthomodular identity
\eqref{e:ModularIdentity2}, and for which if $a_i\in \mathcal{P}$
and $a_i\bot a_j$ ($i\neq j$), this implies that $\bigvee a_i$
exists for $i=1,2,\ldots$. Remark that event structures (or
$\sigma$-orthocomplete orthomodular posets) need not to be
lattices. However, lattices are very general structures  and
encompass most important examples. Consequently,  we will work
with orthomodular lattices in this paper (and indicate which
results can be easily extended to $\sigma$-orthocomplete
orthomodular posets).

\nd There are other general approaches to statistical theories. One
of them is the convex operational one
\cite{wilce,Barnum-Wilce-2006,Barnum-Wilce-2009,Barnum-Wilce-2010},
which consists on imposing axioms on a convex structure (formed by
physical states). Indeed, the convex operational approach is even
more general than the quantum logical, but we will not discuss this
issue in detail here (although a link with it will be discussed in
Section \ref{s:GeneralMethod}).

\section{Cox vs. Kolmogorov}\label{s:CoxReview}

\nd In this Section we will review two different approaches to
probability theory. On one hand, the Cox's approach, in which
probabilities are considered as measures of the plausibility of a
given event or happening. On the other hand, the traditional
Kolmogorovian one, a set theoretical approach which is compatible
with the interpretation of probabilities as frequencies.

\subsection{Kolmogorov}\label{s:Kolmogorov}

\noindent Given a set $\Omega$, let us consider a $\sigma$-algebra
$\Sigma$ of $\Omega$. Then, a probability measure will be given by a
function $\mu$ such that

\begin{subequations}\label{e:kolmogorovian}
\begin{equation}
\mu:\Sigma\rightarrow[0,1]
\end{equation}
\noindent which satisfies
\begin{equation}
\mu(\emptyset)=0
\end{equation}
\begin{equation}
\mu(A^{c})=1-\mu(A),
\end{equation}

\noindent where $(\ldots)^{c}$ means set-theoretical-complement and
for any pairwise disjoint denumerable family $\{A_{i}\}_{i\in I}$

\begin{equation}
\mu(\bigcup_{i\in I}A_{i})=\sum_{i}\mu(A_{i})
\end{equation}

\end{subequations}

\noindent where conditions (\ref{e:kolmogorovian}) are the well
known axioms of Kolmogorov. The triad $(\Omega,\Sigma,\mu)$ is
called a \emph{probability space}. Depending on the context,
probability spaces obeying Eqs. \eqref{e:kolmogorovian} are usually
referred as Kolmogorovian, classical, commutative or boolean
probabilities \cite{Gudder-StatisticalMethods}.

\nd It is possible to show that if $(\Omega,\Sigma,\mu)$ is a
kolmogorovian probability space, the \emph{inclusion-exclusion
principle} holds

\begin{equation}\label{e:SumRule}
\mu(A\cup B)=\mu(A)+\mu(B)-\mu(A\cap B)
\end{equation}

\noindent or (as expressed in logical terms)

\begin{equation}\label{e:SumRuleLogical}
\mu(A \vee B)=\mu(A)+\mu(B)-\mu(A \wedge B)
\end{equation}

\noindent As remarked in \cite{Redei99}, Eq. \eqref{e:SumRule} was
considered as crucial by von Neumann for the interpretation of
$\mu(A)$ and $\mu(B)$ as relative frequencies. If $N_{(A\cup B)}$,
$N_{(A)}$, $N_{(B)}$, $N_{(A\cap B)}$ are the number of times of
each event to occur in a series of $N$ repetitions, then
\eqref{e:SumRule} trivially holds.

\nd As we shall discuss below, this principle does no longer hold in
QM, a fact linked to the non-boolean QM-character. Thus, the
relative-frequencies' interpretation of quantum probabilities
becomes problematic \cite{Redei99}. The QM example shows that
non-distributive propositional structures play an important role in
probability theories {\it different from} that of Kolmogorov.

\subsection{Cox's approach}

\nd Propositions of classical logic can be endowed with a Boolean
lattice structure \cite{Boole}. The logical implication
``$\longrightarrow$" is associated with a partial order relation
``$\leq$", the conjunction ``and" with the greatest lower bound
``$\wedge$", disjunction ``or" with the lowest upper bound
``$\vee$", and negation ``not" is associated with complement
``$\neg$". Boolean lattices can be characterized as ortholattices
satisfying:

\begin{itemize}\label{e:equationsboolean}
\item L1. $x\vee x = x$, $x\wedge x = x$ (idempotence)
\item L2. $x\vee y = y\vee x$, $x\wedge y = y\wedge x$ (commutativity)
\item L3. $x\vee (y\vee z) = (x\vee y)\vee z$, $x\wedge (y\wedge z) = (x\wedge y)\wedge z$ (associativity)
\item L4. $x\vee (x\wedge y) = x\wedge (x\vee y) = x$ (absortion)
\item D1. $x\wedge (y\vee z) = (x\wedge y) \vee (x\wedge z)$ (distributivity 1)
\item D2. $x\vee (y\wedge z) = (x\vee y) \wedge (x\vee z)$ (distributivity 2)
\end{itemize}

\nd It is well known that boolean lattices can be represented as
subsets of a given set, with ``$\leq$" represented as set
theoretical inclusion $\subseteq$, ``$\vee$" represented as set
theoretical union ``$\cup$", ``$\wedge$" represented as set
intersection ``$\cap$", and $\neg$ represented as the set
theoretical complement ``$(\ldots)^{c}$".

\vskip 3mm

\nd As a typical feature, Cox develops classical probability theory
as an inferential calculus on boolean lattices. A real valued
function $\varphi$ representing the degree to which a proposition
$y$ implies another proposition $x$ is postulated, and its
properties deduced from the algebraic properties of the boolean
lattice (Eqns. \eqref{e:ComplementationAxioms} and
\eqref{e:equationsboolean}). These algebraic properties define
functional equations \cite{aczel-book} which determine the possible
elections of $\varphi$ up to rescaling. It turns out that
$\varphi(x|y)$ --if suitably normalized-- satisfies all the
properties of a Kolmogorovian probability (Eqs.
\eqref{e:kolmogorovian}). The deduction will be omitted here, and
the reader is referred to
\cite{CoxPaper,CoxLibro,Knuth-2004a,Knuth-2004b,Knuth-2005b} for
detailed expositions.

\vskip 3mm

\nd Despite their formal equivalence, there is a great conceptual
difference between the approaches of Kolmogorov and Cox. In the
Kolmogorovian approach probabilities are naturally interpreted (but
not necessarily) as relative frequencies in a sample space. On the
other hand, the approach developed by Cox, considers probabilities
as a measure of the degree of belief of an intelligent agent, on the
truth of proposition $x$ if it is known that $y$ is true. This
measure is given by the real number $\varphi(x|y)$, and in this way
the Cox's approach is more compatible with a \emph{Bayesian}
interpretation of probability theory.

\section{Quantum vs. classical probabilities}\label{s:probabilities}

\nd In this Section we will introduce quantum probabilities and
look at their differences with  classical ones. Great part of the
hardship faced by  Birkhoff and von Neumann in developing the
logic of quantum mechanics were due to the inadequacies of
classical probability theory. Their point of view was that any
statistical physical theory could be regarded as a probability
theory, founded on a calculus of events. These events should be
the experimentally verifiable propositions of the theory, and the
structure of this calculus was to be deduced from empirical
considerations, which, for the quantum case, resulted in an
orthomodular lattice \cite{BvN,Srinivas}. We remark on the great
generality of this conception: there is no need of restricting it
to physics. Any statistical theory  formulated  as an event
structure fits into this scheme.

\nd In the formulation of both classical and quantum statistical
theories, states can be regarded as representing consistent
probability assignments \cite{mackey-book,wilce}. In the quantum
mechanics instance {\it this ``states as mappings" visualization} is
achieved via \emph{postulating} a function \cite{mikloredeilibro}

\begin{subequations}\label{e:nonkolmogorov}
\begin{equation}
s:\mathcal{P}(\mathcal{H})\rightarrow [0;1]
\end{equation}
\noindent such that:
\begin{equation}\label{e:Qprobability1}
s(\textbf{0})=0 \,\, (\textbf{0}\,\, \mbox{is the null subspace}).
\end{equation}
\begin{equation}\label{e:Qprobability2}
s(P^{\bot})=1-s(P),\end{equation} \noindent and, for a denumerable
and pairwise orthogonal family of projections ${P_{j}}$
\begin{equation}\label{e:Qprobability3}
s(\sum_{j}P_{j})=\sum_{j}s(P_{j}).
\end{equation}
\end{subequations}

\noindent Gleason's theorem
\cite{Gleason,Gleason-Dvurechenski-2009}, tell us that if the
dimension of $\mathcal{H}\geq 3$, any measure $s$ satisfying
\eqref{e:nonkolmogorov} can be put in correspondence with a trace
class operator (of trace one) $\rho_{s}$ via the correspondence:

\begin{equation}\label{e:bornrule2}
s(P):=\mbox{tr}(\rho_{s} P)
\end{equation}

\nd And vice versa: using equation \eqref{e:bornrule2} any trace
class operator of trace one defines a measure as in
\eqref{e:nonkolmogorov}. Thus, equations \eqref{e:nonkolmogorov}
define a probability: to any elementary test (or event), represented
by a projection operator $P$, $s(P)$ gives us the probability that
the event $P$ occurs, and this is experimentally granted by the
validity of Born's rule. But in fact, \eqref{e:nonkolmogorov} is not
a classical probability, because it does not obeys Kolmogorov's
axioms \eqref{e:kolmogorovian}. The main difference comes from the
fact that the $\sigma$-algebra in (\ref{e:kolmogorovian}) is
boolean, while $\mathcal{P}(\mathcal{H})$ is not. Thus, quantum
probabilities are also called non-kolmogorovian (or non-boolean)
probability measures. The crucial fact is that, in the quantum case,
{\it we do not have a $\sigma$-algebra, but an orthomodular lattice
of projections.}

\nd One of the most important ensuing differences expresses itself
in the fact that Eq. \eqref{e:SumRule} is no longer valid in QM.
Indeed, it may happen that

\begin{equation}
s(A)+s(B)\leq s(A\vee B)
\end{equation}

\nd for $A$ and $B$  suitably chosen elementary sharp tests (see
\cite{Gudder-StatisticalMethods}, Chapter 2). Another important
difference comes from the difficulties which appear when one tries
to define a quantum conditional probability (see for example
\cite{Gudder-StatisticalMethods} and \cite{Redei-Summers2006} for a
comparison between classical and quantum probabilities). Quantum
probabilities may also be considered as a generalization of
classical probability theory: while in an arbitrary statistical
theory a state will be a normalized measure over a suitable
$C^{\ast}$-algebra, the classical case is recovered when the algebra
is \emph{commutative}
\cite{Gudder-StatisticalMethods,Redei-Summers2006}.

\nd We are thus faced with the following fact: on the one hand,
there exists a generalization of classical probability theory to
non-boolean operational structures. On the other hand, Cox derives
classical probabilities from the algebraic properties of classical
logic. As we shall see in detail below, this readily implies that
probabilities in CM are determined by the operational structure of
classical propositions (given by subsets of phase space). The
question is: is it possible to generalize Cox's method to arbitrary
propositional structures (representing the operational propositions
of an arbitrary theory) even when they are not boolean? What would
we expect to find? We will see that the answer to the first question
is \emph{yes}, and for the second, it is reasonable to recover
quantum probabilities (Eq. \eqref{e:nonkolmogorov}). This approach
may serve as a solution for a problem posed by von Neumann. In his
words:

\begin{quote}
``In order to have probability all you need is a concept of all
angles, I mean, other than 90. Now it is perfectly quite true that
in geometry, as soon as you can define the right angle, you can
define all angles. Another way to put it is that if you take the
case of an orthogonal space, those mappings of this space on itself,
which leave orthogonality intact, lives all angles intact, in other
words, in those systems which can be used as models of the logical
background for quantum theory, it is true that as soon as all the
ordinary concepts of logic are fixed under some isomorphic
transformation, all of probability theory is already fixed... This
means however, that one has a formal mechanism in which, logics and
probability theory arise simultaneously and are derived
simultaneously.\cite{Redei99}''
\end{quote}
\noindent and, as remarked by M. Redei \cite{Redei99}:

\begin{quote}
``It was simultaneous emergence and mutual determination of
probability and logic what von Neumann found intriguing and not at
all well understood. He very much wanted to have a detailed
axiomatic study of this phenomenon because he hoped that it would
shed ``... a great deal of new light on logics and probability alter
the whole formal structure of logics considerably, if one succeeds
in deriving this system from first principles, in other words from a
suitable set of axioms."(quote) He emphasized --and this was his
last thought in his address-- that it was an entirely open problem
whether/how such an axiomatic derivation can be carried out.''
\end{quote}

\noindent The problem posed above has remained thus far
unanswered, and this work may be considered as concrete step
towards its solution. Before entering  the subject, let us first
review an alternative approach.

\section{Alternative derivation of Feynman's rules}\label{s:KnuthGoyal}

\noindent Refs. \cite{Symmetry}, \cite{GoyalKnuthSkilling}, and
\cite{KnuthSkilling-2012} present a novel derivation of Feynman's
rules for quantum mechanics, based on a modern reformulation
\cite{Knuth-2005b} of Cox's ideas on the foundations of probability
\cite{CoxPaper,CoxLibro}. To start with, an experimental logic of
processes is defined for quantum systems. This is done in such a way
that the resulting algebra is a distributive one. Given $n$
measurements $M_1$,$\ldots$,$M_n$ on a given system, with results
$m_1$, $m_2$, $\ldots$, $m_n$, the later are organized in a
\emph{measuring sequence} $A=[m_1,m_2,\ldots,m_n]$ as a particular
process. The measuring sequence $A=[m_1,m_2,\ldots,m_n]$ must not be
confused with the conditional (logical) proposition of the form
$(m_2,\ldots,m_n|m_1)$. Sequence $A$ has associated a probability
$P(A)=Pr(m_n,\ldots,m_2|m_1)$ of obtaining outcomes $m_2$, $\ldots$,
$m_n$ conditional upon obtaining $m_1$ \cite{Symmetry}.

\noindent If each of the $m_i$'s has two possible values, $1$ and
$2$, a measuring sequence of three measurements is for example
$A_1=[1,2,1]$. Another one could be $A_2=[1,1,2]$, and so on.

\noindent As is explained in Ref. \cite{Symmetry}:

\begin{quotation}
``A particular outcome of a measurement is either atomic or
coarse-grained. An atomic outcome cannot be more finely divided in
the sense that the detector whose output corresponds to the outcome
cannot be sub-divided into smaller detectors whose outputs
correspond to two or more outcomes. A coarse-grained outcome is one
that does not differentiate between two or more outcomes."
\end{quotation}

\noindent Thus, if we want to ``coarse grain" a certain measurement,
say $M_2$, we can unite the two outcomes in a joint outcome $(1,2)$,
yielding the experiment (measurement) $\widetilde{M}_2$. Thus, a
possible sequence obtained by the replacement of $M_2$ by
$\widetilde{M}_2$ could be $[1,(1,2),1]$. This is used to define a
logical operation

\begin{equation}
[m_1,\ldots,(m_i,m'_i),\ldots,m_n]=[m_1,\ldots,m_i,\ldots,m_n]\vee[m_1,\ldots,m'_i,\ldots,m_n]
\end{equation}

\noindent It is intended that sequences of measurements can be
compounded. For example, if we have $[m_1,m_2]$ and $[m_2,m_3]$, we
have also the sequence $[m_1,m_2,m_3]$, paving the way for  the
general definition

\begin{equation}
[m_1,\ldots,m_j,\ldots,m_n]=[m_1,\ldots,m_j]\cdot[m_j,\ldots,m_n]
\end{equation}

\noindent Given measuring sequences $A$, $B$ and $C$, these
operations satisfy

\begin{subequations}\label{e:ExperimentalLogicKnuth}

\begin{equation}
A\vee B=B\vee A
\end{equation}

\begin{equation}
(A\vee B)\vee C=A\vee(B\vee C)
\end{equation}

\begin{equation}
(A\cdot B)\cdot C=A\cdot(B\cdot C)
\end{equation}

\begin{equation}
(A\vee B)\cdot C=(A\cdot C)\vee(B\cdot C)
\end{equation}

\begin{equation}
C\cdot(A\vee B)=(C\cdot A)\vee(C\cdot B),
\end{equation}

\end{subequations}

\noindent and thus, we have commutativity and associativity of the
operation ``$\vee$", associativity of the operation ``$\cdot$", and
right- and left-distributivity of ``$\cdot$" over ``$\vee$".

\noindent We had already seen in Section \ref{s:CoxReview} that the
method of Cox consists of deriving probability and entropy from the
symmetries of a boolean lattice, intended to represent our
propositions about the world, while probability is interpreted as a
measure of knowledge about an inference calculus. Once equations
(\ref{e:ExperimentalLogicKnuth}) are cast, the set-up for the
derivation of Feynman's rules is ready. The path to follow now is to
apply Cox's method to the symmetries defined by equations
(\ref{e:ExperimentalLogicKnuth}). But this cannot be done
straightforwardly. In order to proceed, an important assumption has
to be made: each measuring sequence will be represented by a pair of
real numbers. This -non operational- assumption is justified in
\cite{Symmetry} using Bohr's complementarity principle. As we shall
se below, the method proposed in this article is an alternative one,
which is more direct and systematic, and makes the introduction of
these assumptions somewhat clearer.

\noindent Once a pair of real numbers is assigned to any measurng
sequence, the authors of \cite{Symmetry} reasonably assume that
equations (\ref{e:ExperimentalLogicKnuth}) induce operations onto
pairs of reals numbers. If measuring sequences $A$, $B$, etc. induce
pairs of real numbers $\mathbf{a}$, $\mathbf{b}$, etc., then, we
should have

\begin{subequations}\label{e:ExperimentalLogicComplex}

\begin{equation}
\mathbf{a}\vee \mathbf{b}=\mathbf{b}\vee \mathbf{a}
\end{equation}

\begin{equation}
(\mathbf{a}\vee \mathbf{b})\vee
\mathbf{c}=\mathbf{a}\vee(\mathbf{b}\vee \mathbf{c})
\end{equation}

\begin{equation}
(\mathbf{a}\cdot
\mathbf{b})\cdot\mathbf{c}=\mathbf{a}\cdot(\mathbf{b}\cdot
\mathbf{c})
\end{equation}

\begin{equation}
(\mathbf{a}\vee \mathbf{b})\cdot \mathbf{c}=(\mathbf{a}\cdot
\mathbf{c})\vee(\mathbf{b}\cdot \mathbf{c})
\end{equation}

\begin{equation}
\mathbf{c}\cdot(\mathbf{a}\vee \mathbf{b})=(\mathbf{c}\cdot
\mathbf{a})\vee(\mathbf{c}\cdot \mathbf{b})
\end{equation}

\end{subequations}

\noindent We easily recognize in (\ref{e:ExperimentalLogicComplex})
operations satisfied by the complex numbers' field (provided that
the operations are interpreted as sum and product of complex
numbers). If they constituted the only possible instance, sequences
represented by pairs of real numbers would be complex numbers, and
thus, we could easily have Feyman's rules. However, complex numbers
are not the only entities that satisfy
(\ref{e:ExperimentalLogicComplex}). There are other such entities,
and thus, extra assumptions have to be made in order to restrict
possibilities. These additional assumptions are presented in
\cite{Symmetry} and \cite{GoyalKnuthSkilling}, and improved upon in
\cite{KnuthSkilling-2012}. We list them below (and refer the reader
to Refs. \cite{Symmetry}, \cite{GoyalKnuthSkilling}, and
\cite{KnuthSkilling-2012} for details).

\begin{itemize}

\item Pair symmetry

\item Additivity condition

\item Symmetric bias condition

\end{itemize}

\noindent Leaving aside the fact that these extra assumptions are
more or less reasonable (justifications for their use are given in
\cite{KnuthSkilling-2012}), it is clear that the derivation is quite
indirect: the experimental logic is thus defined in order to yield
algebraic rules compatible with complex multiplication (and the rest
of the strategy is to make further assumptions in order to discard
other fields different from that of  complex numbers). Further, the
experimental logic characterized by equations
(\ref{e:ExperimentalLogicKnuth}) is not the only possibility, as we
have seen in Section \ref{s:LogicOperational}.

\noindent \fbox{\parbox{0.97\linewidth}{In the rest of this work, we
will apply Cox's method to general propositional structures
according to the quantum logical approach. We will see that this
allows for a new perspective which sheds light onto the structure of
non-boolean probabilities, and is at the same time susceptible of
great generalization. It opens the door to a general derivation of
alternative kinds of probabilities, including quantum and classical
theories as particular cases.}}

\noindent Yet another important remark is in order. As noted in
the Introduction, the work presented in \cite{Symmetry},
\cite{Knuth-2005b}, and \cite{GoyalKnuthSkilling} -as well as
ours- is a combination of two approaches: 1) the one which defines
propositions in an empirical way (something which it shares with
the OQL approach) and 2) that of Cox. Cox's spirit was to derive
probabilities out of  Chomsky's generative propositional
structures that are ingrained in our brain \cite{chom}, and this
boolean structure is independent of any experimental information.
This does not imply, though, that the empirical logic needs to
satisfy the same algebra than pervades our thinking, and that is
indeed what happens. In this sense, any derivation  involving
empirical or operational logics deviates from the original intent
of Cox. As we shall see, this is not a problem, but rather an
important advantage in practice.

\section{Cox's method applied to non-boolean
algebras}\label{s:CoxNon-Boolean}

\noindent As  seen in Section \ref{s:LogicOperational}, {\it
operationally motivated axioms imposed on a lattice's propositional
structure can be used to describe quantum mechanics and other
theories as well.} Disregarding the discussion about the operational
validity of this construction, we are only interested in the fact
that the embodiment is feasible. Similar constructions can be made
for many physical systems, beyond quantum mechanics: the connection
between any theory and experience is given by an event structure
(elementary tests), and these events can be organized in a lattice
structure in most examples of interest.

\nd {\sf Thus, our point of departure will be the fact that physical
systems can be represented by propositional lattices, and that these
lattices need not be necessarily distributive. We will consider
atomic orthomodular lattices. \noindent Given a system $S$, and its
propositional lattice $\mathcal{L}$, we proceed to apply Cox's
method in order to develop an inferential calculus on
$\mathcal{L}$.}

\subsection{Classical Mechanics}\label{s:ClassicalDerivation}

\nd   We  start with classical mechanics (that theory satisfying
Hamilton's equations). Given a classical system $S_{C}$, the
propositional structure is a boolean one, isomorphic to a perfectly
boolean lattice used in our logical language (i.e., regarding its
algebraic structure, it is the same as the one used by Cox).
Accordingly, as shown in Section \ref{s:CoxReview} (proceeding in
the same way as Cox \cite{CoxLibro,CoxPaper}), the corresponding
probability calculus has to be the one which obeys the laws of
Kolmogorov (it satisfies -in particular- equations
\ref{e:kolmogorovian}), and the corresponding information measure is
Shannon's, as expected.

\subsection{Quantum case}\label{s:QMDerivation}

\nd   As shown by Birkoff and von Neumann in \cite{BvN}, if we
follow the above path and try to define the propositional
structure for a quantum system $S_{Q}$ we find an orthomodular
lattice $\mathcal{L}_{v\mathcal{N}}(\mathcal{H})$ isomorphic to
the lattice of projections $\mathcal{P}(\mathcal{H})$. What are we
going to find if we apply instead Cox's method? It stands to
reason that we would encounter a non-boolean probability measure
with the properties \emph{postulated} in Section
\ref{s:probabilities} (Eqs. \eqref{e:nonkolmogorov}). Let us see
that this is indeed the case.

\nd The first thing to remark is that in this derivation we assume
to have a non-boolean lattice
$\mathcal{L}_{v\mathcal{N}}(\mathcal{H})$, isomorphic to the
lattice of projections $\mathcal{P}(\mathcal{H})$. We must show
that the ``degree of implication" measure $s(\cdots)$ demanded by
Cox's method  satisfies Eqs. \eqref{e:nonkolmogorov}. We will only
consider the case of prior probabilities. This  means that we ask
for the probability that a certain event happens for a given state
of affairs, i.e., a concrete preparation of the system under
certain circumstances (which could be natural or artificial).
Thus, we are looking for a function to the real numbers $s$ such
that it is non-negative and $s(P)\leq s(Q)$ whenever $P\leq Q$.

\nd Under these assumptions, let us consider the operation
``$\vee$". As the direct sum of subspaces is associative,
``$\vee$" will be associative too. If $P$ and $Q$ are orthogonal
projections ($P\perp Q$), then $P\wedge Q=\mathbf{0}$ (otherwise,
there would be a vector in $P$ which is not orthogonal to every
vector of $Q$). Next, we consider the relationship between $s(P)$,
$s(Q),$ and $s(P\vee Q)$. As $P\wedge Q=\mathbf{0}$, it should
happen that

\begin{equation}\label{e:s(or)}
s(P\vee Q)=F(s(P),s(Q)),
\end{equation}

\noindent with $F$ a function to be determined. Add now a third
proposition $R$ (notice that, for doing this, we need a space of
dimension $d\geq 3$, an interesting analogy with Gleason's
theorem), such that $P\perp R$, $Q\perp R,$ and $Q\perp P$ (and
thus $P\wedge R=\mathbf{0}$, $Q\wedge R=\mathbf{0},$ and $Q\wedge
P=\mathbf{0}$). Build now the element $(P\vee Q)\vee R$. Then,
because of the associativity of ``$\vee$", we arrive at the
following result

\begin{equation}
s((P\vee Q)\vee R)=s(P\vee(Q\vee R)),
\end{equation}

\nd and thus (using \eqref{e:s(or)}),

\begin{equation}\label{e:FunctEq1}
F(F(s(P),s(Q)),s(R))=F(s(P),F(s(Q),s(R))).
\end{equation}

\noindent The algebraic properties of associativity for $\vee$ and
$\perp$ are the only prerequisite for this result. Thus, proceeding
as in \cite{Knuth-2004a,Knuth-2004b,Knuth-2005b} (and using the
solutions to functional equations of the form \eqref{e:FunctEq1}
studied in \cite{aczel-book}), we have that --up to a re-scaling:

\begin{equation}
s(P\vee Q)=s(P)+s(Q).
\end{equation}

\nd whenever $P\perp Q$. It thus follows that for any finite family
of orthogonal projections $P_j$, $1\leq j\leq n$, we have $s(P_1\vee
P_2\vee\cdots\vee P_n)=s(P_1)+s(P_2)+\cdots+s(P_n)$. Now, as any
projection $P$ satisfies $P\leq \mathbf{1}$, then $s(P)\leq
s(\mathbf{1})$, and we can assume without loss of generality the
normalization condition $s(\mathbf{1})=1$. Thus, for any denumerable
pairwise orthogonal infinite family of projections $P_j$, we have
for each $n$

\begin{equation}
\sum_{j=1}^{n}s(P_j)=s(\bigvee_{j=1}^{n}P_j)\leq 1.
\end{equation}

\nd As $s(P_j)\geq 0$ for each $j$, the sequence
$s_n=s(\bigvee_{j=1}^{n}P_j)$ is monotone, bounded from above, and
thus converges. We  write then

\begin{equation}
s(\bigvee_{j=1}^{\infty}P_j)=\sum_{j=1}^{\infty}s(P_j),
\end{equation}
\nd and we  recover condition \eqref{e:Qprobability3} of the
axioms of quantum probability. Now, given any proposition
$\mathcal{L}_{v\mathcal{N}}(\mathcal{H})$, consider $P^{\perp}$.
As $P\vee P^{\perp}=\mathbf{1}$, and $P$ is orthogonal to
$P^{\perp}$, we  have

\begin{equation}
s(P\vee P^{\perp})=s(P)+s(P^{\perp})=s(\mathbf{1})=1.
\end{equation}
\nd In other words,

\begin{equation}
s(P^{\perp})=1-s(P),
\end{equation}
\nd which is nothing but condition \eqref{e:Qprobability2}. On the
other hand, as $\mathbf{0}=\mathbf{0}\vee\mathbf{0}$ and
$\mathbf{0}\bot\mathbf{0}$, then
$s(\mathbf{0})=s(\mathbf{0})+s(\mathbf{0})$, and thus,
$s(\mathbf{0})=0$, which is condition \eqref{e:Qprobability1}.

\noindent \fbox{\parbox{0.97\linewidth}{ \emph{This Section shows
that using the algebraic properties of $\mathcal{L}_{v\mathcal{N}}$,
it is possible to derive the form of the quantum probabilities
which, on the light of this discussion, do not need to be
postulated.}}}\vskip 2mm

\nd Thus, we have proved that $s$ is a probability measure on
$\mathcal{L}_{v\mathcal{N}}$. Is there any possibility that $s$
differs from the standard formulation of a quantum probability
measure as a density matrix using the Born's rule? The answer is no,
and this is granted by Gleason's theorem, because we have proved
that $s$ satisfies Eqs. \eqref{e:nonkolmogorov}, and Gleason's
theorem leaves no alternative (if the dimension of $\mathcal{H}\geq
3$).

\nd An important question is the following: which will be the
effect of non-distributivity? As we saw in Section
\ref{s:probabilities}, classical probabilities are sub-additive,
i.e., they satisfy

\begin{equation}\label{e:SubadditiveProb}
\mu(A\vee B)\leq\mu(A)+\mu(B),
\end{equation}

\noindent and this is linked to the stronger assertion of Eq.
\eqref{e:SumRule} (see also \cite{mikloredeilibro}, page $104$). But
as we have seen, it is indeed the case that the analogous of Eq.
\eqref{e:SubadditiveProb} does not hold for quantum probabilities.

\vskip 3mm

\nd We show below that this derivation is susceptible of
generalization. Indeed, the derivation relies mainly on the
algebraic properties of the lattice of projections, i.e., in its
\emph{non-distributive} lattice structure.

\subsection{A Finite Non-distributive Example}\label{s:FiniteCase}

\begin{figure}\label{f:Ejemplo1}

\begin{center}

\includegraphics[width=8cm]{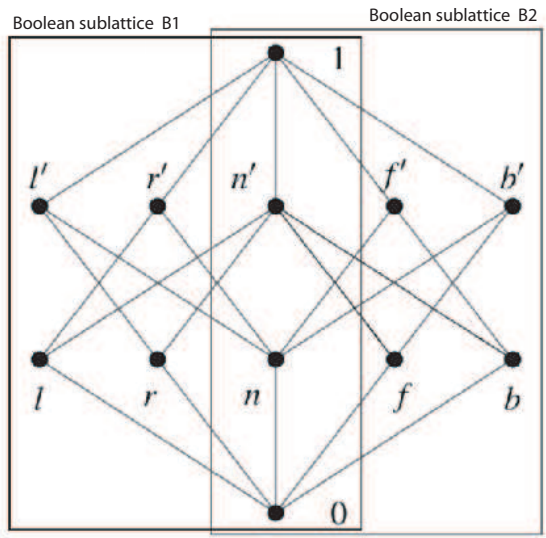}

\caption{\small{Hasse diagram for $L_{12}$. The boolean sublattices
B1 and B2 corresponding to the two complementary contexts are
indicated by squared lines.}}
\end{center}

\end{figure}

\nd There are many systems of interest which can be represented by
finite lattices. Many toy models serve to illustrate special
features of different theories. Let us start first by analyzing
$L_{12}$, a non distributive lattice which may be considered as the
union of two incompatible experiments \cite{svozillibro}. The Hasse
diagram of $L_{12}$ is represented in Fig. \ref{f:Ejemplo1}. An
example of $L_{12}$ is provided by a firefly  which flies inside a
room. The first experiment is to test if the firefly shines on the
right side (r) of the room, or on the left (l), or if it does not
shine at all (n). Other experiment consists in testing if the
firefly shines at the front of the room (f), or on the back (b) of
it, or if it does not shine (n). It is forbidden to make both
experiments at the same time (and thus, emulating contextuality).

\nd Applying the Cox's method to the boolean sublattices of $L_{12}$
(see Fig. \ref{f:Ejemplo1}) and suitably normalizing, we obtain
classical probabilities for each one of them. It is also easy to
find that $P(l)+P(r)+P(f)+P(b)+P(n)=2-P(n)$, a quantity which may be
greater than $1$. The last equation implies explicitly that the
exclusion-inclusion law \eqref{e:SumRule} does not hold. This is due
to the global non-distributivity of $L_{12}$. It is also easy to
show that $P(l)+P(r)=P(f)+P(b)$, yielding a non trivial relationship
between atoms (something which does not occur in a distributive
lattice).

\subsection{General Derivation}\label{s:GeneralDerivation}

\nd Let $\mathcal{L}$ be an atomic orthomodular lattice. We will
compute prior probabilities, i.e., we will assume that
$\mathcal{L}$ represents the propositional structure of a given
system --physical or not--, and that we want to ascertain how
likely a given event is when represented by a lattice element
$a\in\mathcal{L}$. One assumes that the system has undergone a
preparation, i.e., we take it that there exists a \emph{state of
affairs}.

\nd We must define a function $s:\mathcal{L}\longrightarrow
\mathbb{R}$, such that it is always non-negative

\begin{subequations}
\begin{equation}
s(a)\geq 0\,\,\forall a\in\mathcal{L}
\end{equation}
and is also order preserving
\begin{equation}
a\leq b \longrightarrow s(a)\leq s(b).
\end{equation}
\end{subequations}

\nd We will show that under these rather general assumptions a
probability theory can be developed. The order preserving assumption
readily implies that $s(a)\leq s(\textbf{1})$ for all
$a\in\mathcal{L}$. We will also assume that $s(\textbf{1})=K$, a
finite real number.

\nd Now, as an ortholattice is complemented (Eqs.
\eqref{e:ComplementationAxioms}), we will always have that
$\neg\neg a=a$ for all $a\in\mathcal{L}$. Accordingly,

\begin{equation}\label{e:Sdenonoa}
s(\neg\neg a)=s(a),
\end{equation}

\nd for all $a$. Next, it is also reasonable to assume that $s(\neg
a)$ is a function of $s(\neg a)$, say $s(\neg a)=g(s(a))$. Thus,
Eqs. \eqref{e:Complement1} and \eqref{e:Sdenonoa} imply

\begin{equation}\label{e:FunctionalEqG}
g(g(s(a)))=s(a),
\end{equation}
\nd or, in other words,

\begin{equation}\label{e:FunctionalEqGbis}
g(g(x))=x,
\end{equation}

\nd for positive $x$. A family of functions which satisfy
\eqref{e:FunctionalEqG} are $g(x)=x$ and $g(x)=c-x$, where $c$ is a
real constant\footnote{There are additional solutions to this equation,
but, if we suitably choose scales, we can disregard them for
these two cases. For a discussion on re-scaling of measures, their
meaning and validity, see \cite{CoxLibro}.}. We discard the first
possibility because if true, we would have $s(\mathbf{0})=s(\neg
\mathbf{1})=g(s(\mathbf{1}))=s(\mathbf{1})$. But if
$s(\mathbf{0})=s(\mathbf{1})$, because of $\mathbf{0}\leq
x\leq\mathbf{1}$ for all $x\in\mathcal{L}$, we have
$s(\mathbf{0})=s(x)=s(\mathbf{1})$, and our measure would be
trivial. Thus, the only non-trivial option ---up to rescaling--- is
$s(\neg a)=c-s(a)$.

\nd Now, let us see what happens with the ``$\vee$" operation. As
$\mathcal{L}$ is orthocomplemented,  the orthogonality notion for
 elements is available (see Section \ref{s:LatticeTheory}). If
$a,b\in\mathcal{L}$ and $a\bot b$, because of \eqref{e:Complement2},
we have that $a\wedge b=\mathbf{0}$. Thus, it is reasonable to
assume that $s(a\vee b)$ is a function of $s(a)$ and $s(b)$ only,
i.e., $s(a\vee b)=f(s(a),s(b))$. By associativity of the ``$\vee$"
operation, $(a\vee b)\vee c=a\vee(b\vee c)$ for any
$a,b,c\in\mathcal{L}$, and this implies  then that $s((a\vee b)\vee
c)=s(a\vee(b\vee c))$. If $a$, $b$, and $c$ are orthogonal, we will
have for the left hand side $s((a\vee b)\vee
c)=f(f(s(a),s(b)),s(c))$ and $s(a\vee (b\vee
c))=f(s(a),f(s(b),s(c)))$ for the right hand side. Thus,

\begin{equation}
f(f(s(a),s(b)),s(c))=f(s(a),f(s(b),s(c))),
\end{equation}

\nd or, in a simpler form

\begin{equation}\label{e:FunctionalEq2}
f(f(x,y),z)=f(x,f(y,z)),
\end{equation}

\nd with $x$, $y,$ and $z$ positive real numbers. As shown in
\cite{aczel-book}, the only solution (up to re-scaling) of
\eqref{e:FunctionalEq2} is $f(x,y)=x+y$. We have thus shown that if
$a\bot b$

\begin{equation}
s(a\vee b)=s(a)+s(b),
\end{equation}
\nd and we will also have

\begin{equation}\label{e:FiniteSum}
s(a_1\vee a_2\cdots\vee a_n)=s(a_1)+s(b_2)+\cdots+s(a_n),
\end{equation}

\nd whenever $a_1$, $a_2$, $\cdots$, $a_n$ are pairwise orthogonal.
Suppose now that $\{a_i\}_{i\in\mathbb{N}}$ is a family of pairwise
orthogonal elements of $\mathcal{L}$. For any finite $n$, we have
that $a_1\vee a_2\vee\cdots\vee a_n\leq \mathbf{1}$, and thus
$s(a_1\vee a_2\vee\cdots\vee a_n)=s(a_1)+s(b_2)+\cdots+s(a_n)\leq
s(\mathbf{1})=K$. Then, $s_n=s(a_1\vee a_2\vee\cdots\vee a_n)$ is a
monotone sequence bounded from above, and thus it converges to a
real number. As
$\bigvee\{a_{i}\}_{i\in\mathbb{N}}=\lim_{n\longrightarrow\infty}\bigvee_{i=1}^{n}a_i$,
we can write

\begin{equation}
s(\bigvee\{a_i\}_{i\in\mathbb{N}})=\sum_{i=1}^{\infty}s(a_i).
\end{equation}

\nd In any orthomodular lattice we have $\mathbf{1}\bot \mathbf{0}$
(because $\mathbf{0}\leq \neg \mathbf{1}=\mathbf{0}$), and
$\mathbf{1}\vee\mathbf{0}=\mathbf{1}$. Thus, $s(\mathbf{1}\vee
\mathbf{0})=s(\mathbf{1})=s(\mathbf{1})+s(\mathbf{0})$. Accordingly,
$s(\mathbf{0})=0$. As $\neg \mathbf{1}=\mathbf{0}$,
$s(\mathbf{0})=c-s(\mathbf{1})$. Thus, $s(\mathbf{1})=c$ and then,
$c=K$. We will not lose generality if we assume the normalization
condition $K=1$.

\vskip3mm

\nd The results of this section show that in \emph{any} orthomodular
lattice, a reasonable measure $s$ of plausibility of a given event
must satisfy that, for any orthogonal denumerable family
$\{a_i\}_{i\in\mathbb{N}}$, one has (up to rescaling)

\begin{subequations}\label{e:GeneralizedProbability}
\begin{equation}
s(\bigvee\{a_i\}_{i\in\mathbb{N}})=\sum_{i=1}^{\infty}s(a_i)
\end{equation}
\begin{equation}
s(\neg a)= 1-s(a)
\end{equation}
\begin{equation}
s(\mathbf{0})=0.
\end{equation}
\end{subequations}

\vskip3mm

\nd Why do Eqs. \eqref{e:GeneralizedProbability} define
non-classical (non-Kolmogorovian) probability measures? In a
non-distributive orthomodular lattice there always exist elements
$a$ and $b$ such that

\begin{equation}
(a\wedge b)\vee (a\wedge\neg b)<a,
\end{equation}
\nd so that (using $(a\wedge\neg b)\bot(a\wedge b$)),
$s((a\wedge\neg b)\vee(a\wedge b))=s(a\wedge\neg b)+s(a\wedge b)\leq
s(a)$. The inequality can be strict, as the quantum case shows. But
in any classical probability theory, by virtue of the
inclusion-exclusion principle (Eqn. \eqref{e:SumRuleLogical}), we
always have $s(a\wedge\neg b)+s(a\wedge b)= s(a)$. This simple fact
shows that our measures will be non-classical in the general case.

\vskip 3mm

\noindent \fbox{\parbox{0.97\linewidth}{ In this way, we provide an
answer to the problem posed by von Neumann and discussed in Section
\ref{s:probabilities}: \emph{the algebraic and logical properties of
the operational event structure determine up to rescaling the
general form of the probability measures which can be defined over
the lattice.} Accordingly, we did present a generalization of the
Cox method to non-boolean structures, namely orthomodular lattices,
and then \emph{we have indeed  deduced a generalized probability
theory.}}}\vskip 2mm

\nd Remark that boolean lattices are also orthomodular. This means
that our derivation is a generalization of that of Cox, and when we
face a boolean structure, classical probability theory will be
deduced exactly as in \cite{CoxPaper,CoxLibro}.

\noindent Another important remark is that -as the reader may check-
the derivation presented here \emph{is also valid} for
$\sigma$-orthocomplete orthomodular posets (see Section
\ref{s:LogicOperational} of this work), and thus, of great
generality. Indeed, in these structures the disjunction exists for
denumerable collections of orthogonal elements and it is
associative. This fact, together with orthocomplementation, make the
above results remain valid in these structures. But as it was
mentioned in Section \ref{s:LogicOperational},
$\sigma$-orthocomplete orthomodular posets are not lattices in the
general case. In this way, we remark that the generalization of the
Cox's method goes well beyond lattice theory. Thus, yielding a
generalized probability calculus which includes the general physical
approach presented for example, in \cite{Gudder-StatisticalMethods}.

\subsection{A general methodology}\label{s:GeneralMethod}

\begin{figure}\label{f:MetodoGeneral}

\begin{center}

\includegraphics[width=8cm]{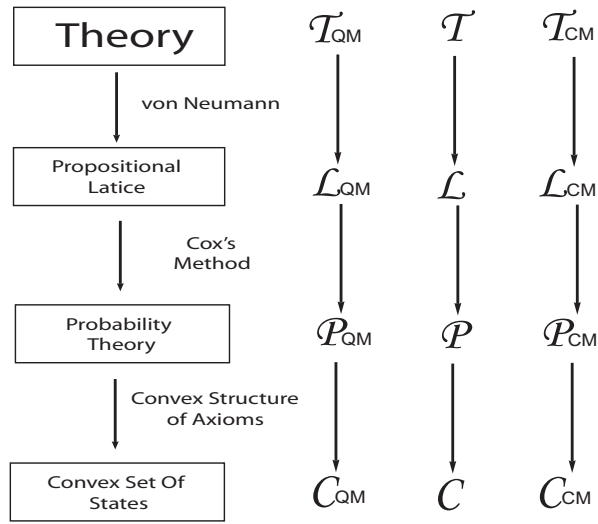}

\caption{\small{Schematic representation of the method proposed
here. A general theory $\mathcal{T}$ determines via the von Neumann
approach the algebraic structure of the set of elementary tests.
Then, by applying Cox's method, it is  possible to determine the
general properties of the canonical probability theory assigned to
$\mathcal{T}$. Next, by assigning particular values to prior
probabilities of atoms, all states will be determined and thus, the
form of $\mathcal{C}$, the convex set of states. The quantum
mechanical ($QM$) and the classical ($CM$) case are shown as the
extreme instances of a vast family of theories.}}
\end{center}
\end{figure}

\nd At this stage it is easy to envisage just how a general method
could be developed. One first starts  by identifying the algebraic
structure of the elementary tests of a given theory $\mathcal{T}$.
They determine all observable quantities and, in the most important
examples, they are endowed with a lattice structure. Once the
algebraic properties of the pertinent lattice are fixed, Cox's
method can be used to \emph{determine} (up to rescaling) the general
properties of prior probabilities. Of course, it does not
predetermine the particular values that these probabilities might
take on the atoms of the theory. Such an specification would amount
to determine a \emph{particular state} of the system under scrutiny.
The whole set of these states can be rearranged to yield a convex
set. The path we have followed here is illustrated in Fig.
\ref{f:MetodoGeneral}, with the classical and quantum cases as
extreme examples of a vast family.

\nd The method described in this work can be seen as a general
epistemological background for a huge family of scientific theories.
In order to have a (quantitative) scientific theory, we must be able
to make predictions on certain events of interest. Events are
regarded here as propositions susceptible of being tested. Thus, one
starts from an inferential calculus which allows for  quantifying
the degree of belief on the certainty of an event $x$, if it is
known that event $y$ has occurred. The crucial point is that event
structures are not always organized as boolean lattices (QM being
perhaps the most spectacular example). Thus, in order to determine
the general properties of the probabilities of a given theory (and
thus all possible states by specifying particular values of prior
probabilities on the atoms), we must apply Cox's method to lattices
more general than boolean ones.

\section{Conclusions}\label{s:Conclusions}
\nd By complementing the results presented in
\cite{Caticha-99,Knuth-2004a,Knuth-2004b,Knuth-2005a,Knuth-2005b,Symmetry,KnuthSkilling-2012},
in this work we showed that it is possible to combine Cox's method
for the foundations of classical probability theory and the OQL
approach, in order \emph{to give an alternative derivation of what
is known as non-kolmogorovian (or non commutative) probability
theory} \cite{Gudder-StatisticalMethods,Redei-Summers2006}.

\vskip 3mm

\noindent Most physical probabilistic theories of interest can be
endowed with orthomodular lattices. The elements of these lattices
represent the events of that theory. As in
\cite{Knuth-2004b,Knuth-2005a}, we have studied how the algebraic
properties of the lattices determine
---up to rescaling--- the general form of the probabilities
associated to the given physical theory. In this way, we provided a
new formulation of the approach to physics based on
non-kolmogorovian probability theory
\cite{Gudder-StatisticalMethods}.

\vskip 3mm

\noindent Differently from
\cite{Caticha-99,Knuth-2004a,Knuth-2004b,Knuth-2005a,Knuth-2005b,Symmetry,KnuthSkilling-2012},
in this work we have focused on the application of the Cox's method
to the von Neumann's lattice of projection operators in a separable
Hilbert space and general orthomodular lattices as well, studying
their particularities. In doing so, we obtained the von Neumann's
axioms of quantum probabilities, and thus (using Gleason's theorem),
quantum probabilities. In this way, our approach includes quantum
mixtures explicitly and naturally.

\vskip 3mm

\noindent It is interesting to remark (as in
\cite{Knuth-2004b,Knuth-2005a}) that this construction is not
necessarily restricted to physics; any probabilistic theory endowed
with a lattice structure of events will follow the same route.

\vskip 3mm

\noindent We have also found that the method can be easily extended
to $\sigma$-orthomodular posets (which are not lattices in the
general case), and thus, it contains generalizations of QM.

\vskip 3mm

\nd In deriving quantum probability out of the lattice properties,
we have shown in a direct way how non-distributivity forbids the
derivation of a Kolmogorovian probability. This sheds light on the
structure of quantum probabilities and on their differences with
the classical case. In particular, we have shown that \emph{the
properties of the underlying algebraic structure imply that the
inclusion-exclusion principle is not valid for the von Neumann's
lattice}. And, moreover, that it can also fail in the more general
framework of orthomodular (non-distributive) lattices. We have
shown the explicit violation of the inclusion-exclusion principle
for the finite example of the Chinese lantern and also for a non
trivial relationship between atoms.

\vskip 3mm

\noindent Furthermore, we have provided a non-trivial connection
between the Cox's approach and the problem posed by von Neumann
regarding the axiomatization of probabilities. This is a novel
perspective on the origin and axiomatization of the probabilities
which appear in quantum theory. We believe that -far from being a
definitive answer to the interpretation of quantum probabilities-
this is an interesting topic to be further investigated.

\vskip 3mm

\nd Using Cox's approach, Shanon's entropy can be deduced as a
natural measure of information over the boolean algebra of classical
propositions \cite{CoxLibro,KnuthSkilling-2012,Knuth-2005b}. We will
provide a detailed study of what happens with orthomodular latices
elsewhere.

\vskip 3mm

\nd The strategy followed in this work suggests that we are at the
gates of a great generalization. The general rule for constructing
probabilities would read as follows:

\begin{itemize}

\item 1 - We start by identifying the operational logic of our physical
system. The characteristics of this ``empirical" logic depends
both on physical properties of the system and on the election of
the properties that we assume in order to study the system. This
can be done in a standard way, and the method is provided by the
OQL approach.

\item 2 - Once the operational logic is identified, the symmetries
of the lattice are used to define the properties of the ``degree
of implication" function, which will turn out to be the
probability function associated to that particular logic. Remark
that the same physical system may have different propositional
structures, depending of the election of the observers. For
example, if we look at the observable ``electron's" charge, we
will face classical propositions, but if wee look at its momentum
and position, we will have a non-boolean lattice.

\end{itemize}

\nd This method is  not only of  physical interest, but also of
mathematical one, because one is solving the problem of
characterizing probability measures over general lattices (and other
structures as well). A final remark: our approach deviates from that
of Cox, in the sense that we look for an empirical logic which would
be intrinsic to the system under study, and because of that, not
only referred to our ignorance about it, but to assumptions about
its nature. But this is not a problem, but an advantage: any problem
which can be transformed into the language of lattice theory (or the
more general framework of $\sigma$-othocomplemented orthomodular
posets) may fall into this scheme, and a probability theory can be
developed using the method described in this work.


\vskip1truecm

\noindent {\bf Acknowledgements} \noindent This work was partially
supported by the  grants PIP N$^o$ 6461/05 amd 1177 (CONICET). Also
by the projects FIS2008-00781/FIS (MICINN) - FEDER (EU) (Spain, EU).
The contribution of F. Holik to this work was done during his
postdoctoral stance (CONICET) at Universidad Nacional de La Plata,
Instituto de F\'{\i}sica (IFLP-CCT-CONICET), C.C. 727, 1900 La
Plata, Argentina.

\end{document}